\documentclass{pasa}%

\usepackage{graphicx}

\DeclareMathOperator{\sinc}{sinc}
\newcommand{\PaperI}{Paper I\nocite{Ord2019}}
\newcommand{\Kron}{\delta_s^0}
\newcommand{\vcstools}{VCSTools}
\newcommand{\psrslowB}{B0950$+$08}
\newcommand{\psrslowJ}{J0953$+$0755}
\newcommand{\psrkaurJ}{J2241$-$5236}
\newcommand{\psrbhatJ}{J0437$-$4715}
\newcommand{\dmunits}{\ensuremath{\text{pc}\,\text{cm}^{-3}}}

\title[MWA tied-array processing III]{MWA tied-array processing III: Microsecond time resolution via a polyphase synthesis filter}

\author[McSweeney et al.]{S. J. McSweeney$^1$, S. M. Ord$^2$, D. Kaur$^1$, N. D. R. Bhat$^1$, B. W. Meyers$^{1,2,3}$, S. E. Tremblay, J. Jones$^1$, B. Crosse$^1$, K. R. Smith$^1$
\affil{$^1$International Centre for Radio Astronomy Research (ICRAR), Curtin University, 1 Turner Avenue, Technology Park, Bentley, 6102, W.A., Australia}
\affil{$^2$CSIRO Astronomy and Space Science, PO Box 76, Epping, NSW 1710, Australia}
\affil{$^3$Department of Physics and Astronomy, University of British Columbia, 6224 Agricultural Road, Vancouver, BC V6T 1Z1, Canada}
}

\jid{PASA}
\doi{10.1017/pas.\the\year.xxx}
\jyear{\the\year}

\usepackage{aas_macros}
\usepackage{hyperref}
\hypersetup{colorlinks,citecolor=blue,linkcolor=blue,urlcolor=blue}

\begin{document}

\begin{frontmatter}
\maketitle

\begin{abstract}
A new high time resolution observing mode for the Murchison Widefield Array (MWA) is described, enabling full polarimetric observations with up to $30.72\,$MHz of bandwidth and a time resolution of $\sim 0.8\,\mu$s.
This mode makes use of a polyphase synthesis filter to ``undo'' the polyphase analysis filter stage of the standard MWA's Voltage Capture System (VCS) observing mode.
Sources of potential error in the reconstruction of the high time resolution data are identified and quantified, with the S/N loss induced by the back-to-back system not exceeding $-0.65\,$dB for typical noise-dominated samples.
The system is further verified by observing three pulsars with known structure on microsecond timescales.
\end{abstract}

\end{frontmatter}

\section{INTRODUCTION}
\label{sec:intro}

Some of the most exciting advances in time-domain astronomy have only been made possible by pushing the capabilities of latest generation telescopes to be sensitive to signals of shorter and shorter duration.
The serendipitous discovery of pulsars in the late 1960's is perhaps the prototypical example \citep{Hewish1968}.
In more recent times, the ongoing effort to detect nanohertz gravitational waves by means of pulsar timing arrays requires the continual monitoring of the times of arrival (TOAs) of millisecond pulsars (MSPs) with microsecond accuracy \citep[e.g.][]{Hobbs2017}.
Pulsars are also known to exhibit temporal structures on microsecond and even nanosecond time scales \citep[e.g.][]{Craft1968,Hankins2003}, providing major clues for the underlying radio emission mechanism \citep[e.g.][]{Cordes1981,Popov2002}.
Similarly, fast radio bursts (FRBs) have been shown to exhibit temporal sub-millisecond structures that either point to the intrinsic emission mechanism or to interesting propagation effects occurring in the intergalactic medium \citep{Farah2018,Hessels2019}.
All of these examples serve to illustrate the scientifically important and still largely untapped parameter space that is only accessible to telescopes equipped with a sufficiently high time resolution observing mode.

The Murchison Widefield Array \citep[MWA;][]{Tingay2013} is a low-frequency ($\sim80$ to $300\,$MHz) aperture array telescope located at the Murchison Radio Observatory (MRO) in Western Australia.
Now in its second phase of development \citep[Phase II;][]{Wayth2018}, it consists of 256 `tiles' (sets of $4\times4$ cross-dipole antennas) distributed over an area approximately $5.3\,$km in diameter, 128 of which can be used at a single time to form an interferometer.
Originally conceived as an imaging telescope (which requires only the time-averaged cross-correlation products of the tiles, or `visibilities', to be retained on disk), it was subsequently augmented with the functionality to capture the raw complex voltages of each tile, known as the Voltage Capture System \citep[VCS;][]{Tremblay2015}.
This system has enabled the MWA to be used as a premier instrument for high-time resolution studies of transient signals, especially pulsars \citep[e.g.][]{Bhat2016,McSweeney2017,Meyers2018,Kirsten2019}.

Although the tile voltages are sampled at a (Nyquist) rate of $655.36\,$MHz, these data undergo several stages of processing before finally being written to disk.
After preliminary filtering and digitisation, the raw voltages are subjected to a two-stage frequency analysis filter, which trades time resolution for increased frequency resolution.
In the MWA's case, both stages of the analysis filter were implemented as polyphase filterbanks \citep[PFBs;][]{Harris2011,Prabu2015}.
The first stage (`coarse') PFB reduces the effecting sampling rate by a factor of 512, resulting in an array of complex-valued samples with $1.28\,$MHz resolution in frequency (`coarse' channels) and $\sim0.8\,\mu$s in time.
In the second stage (`fine') PFB, each coarse channel is further split into $128 \times 10\,$kHz `fine' channels at the cost of decreasing the time resolution to $100\,\mu$s.

In the current MWA system design, only the latter time resolution data product (i.e. $100\,\mu$s) is made available to the user.
While this is sufficient for many pulsar studies \citep[e.g.][]{Oronsaye2015,McSweeney2017,Bhat2018,Meyers2018}, it is nevertheless too coarse for many science applications involving MSPs.
In principle, the original higher time resolution can be recovered from the channelised output (either approximately or exactly) by means of a \textit{synthesis filter}, which acts as an `inverse' operation to the analysis filter.
The conditions under which the original time series can be exactly reproduced depends on the choice of analysis and synthesis filters.

Here, we describe the synthesis filter that is implemented as the (optional) final stage of the tied-array beamforming pipeline, the former stages of which are described in detail in \citet[][hereafter Paper I]{Ord2019} and \citet[][hereafter Paper II]{Xue2019}.
The synthesis filter is applied to the fine-channel output of the beamformer, and recovers the coarse channel time series.
That is, it effectively `undoes' the fine PFB, increasing the available time resolution to $\sim 0.8\,\mu$s.
A brief review of PFBs in general, and their particular implementation in the case of the MWA, are given in Section \ref{sec:pfb}.
The design of the synthesis filter is described in Section \ref{sec:ipfb}, including a discussion of its fidelity, i.e., the appearance of any temporal and spectral artefacts introduced by the synthesis filter itself.
Finally, the practical use of this functionality is demonstrated in Section \ref{sec:pulsardata} through three examples: MWA observations of the PSRs \psrkaurJ{}, \psrbhatJ{}, and \psrslowB{} (\psrslowJ{}).

\section{Polyphase filterbanks (PFB)}
\label{sec:pfb}

PFBs are a type of analysis filter, designed to extract spectral information out of discrete time series data.
They can be considered a generalisation of the more familiar discrete Fourier transform (DFT), and are designed to overcome the undesirably uneven frequency response (i.e. spectral leakage) inherent in the application of DFTs to discretely sampled time series of finite length.
PFBs are well described in standard texts \citep{Crochiere1983,Harris2004,Oppenheim2009}; a clear and concise review of PFBs in the context of radio astronomy is given in \citet{Harris2011}.
Thus, only a brief review of the salient features is given, in order to prepare the reader for the description of the synthesis filter in the following sections.

\subsection{General review and mathematical notation}

The PFB is a transformation from the time domain, $x[n]$, to the frequency domain $X_k[m]$, where the $[\cdot]$ notation denotes a discretised index, $k$ is the channel number, and $n,m \in \mathbb{Z}$ are the time indices for the pre- and post-channelised data, respectively.
Let $K$ be the number of (equally spaced) frequency channels required for some desired spectral resolution.
Although applying a DFT to $N = K$ adjacent time samples in $x[n]$ will produce the desired resolution, the result will be an imperfect representation of the true spectrum because of \textit{spectral leakage}, which is when power that properly `belongs' to some particular frequency bin appears in (or, is \textit{aliased} to) other, nearby bins.
The impossibility of perfectly eliminating spectral leakage can be seen by realising that operating on a finite-length time series is equivalent to multiplying an arbitrarily long time series with a rectangular window function,
\begin{equation}
    w_R[n] = \begin{cases} 1, & 0 \le n < N, \\ 0, & \text{elsewhere}, \end{cases}
\end{equation}
whose effect in the frequency domain is to convolve the ``true'' spectrum of the signal with the Fourier transform of the window function.
In the case of a rectangular window, this is the sinc function.

A common strategy for mitigating spectral leakage is to choose an alternative window function whose Fourier pair is localised in the frequency domain, and which therefore produces a tolerable level of spectral leakage when convolved with the signal's spectrum.
Many possible windowing functions have been identified, which in general trade the `amount' of leakage with the `location' of the leaked components.
For our purposes, it is sufficient to note that the inevitable presence of a windowing function motivates the definition of the \textit{analysis filter}, $h[n]$, and the generalised \textit{windowed DFT}
\begin{equation}
    X_k[m] = \sum_{n=0}^{N-1} h[mM - n]\,x[n]\,e^{-2\pi jkn/K},
    \label{eqn:wdft}
\end{equation}
where $j = \sqrt{-1}$ denotes the imaginary number, $m$ is the time index of the channelised (output) data, and $0 \le k < K$ denotes the channel number.
Eq. \eqref{eqn:wdft} describes the action of performing DFTs on short, windowed segments of the input time series of length $K$.
$M$ is the number of samples that the window is translated along $x[n]$ between successive DFT operations; thus, the index of $h[mM - n]$ represents the shift required in order to produce the spectrum at time $m$.
If $M < K$ then the windows overlap and the resulting channelisation is \textit{oversampled}; if $M = K$, then it is \textit{critically sampled}.
The choice of $h[n]$ is motivated by the shape of its \textit{frequency response} (i.e. its Fourier pair), whose characteristics (e.g. width, location of sidelobes) are chosen according to the advantages they carry in particular contexts.
Leaving $x[n]$ ``unweighted'' is equivalent to choosing $h[n] = w_R[n]$, in which case Eq. \eqref{eqn:wdft} merely describes a DFT performed on each successive window, which in this context is also called a short-time Fourier transform (STFT).
On the other hand, choosing $h[n]$ to be the $\sinc$ function will result in a frequency response that approximates a rectangular window.

It is well known that scaling a function in the time domain produces the inverse scaling in the Fourier domain.
This fact motivates an alternative strategy for mitigating spectral leakage.
Choosing a larger window size, $N = KP$ (for integer $P > 1$), and a corresponding wider analysis filter, will result in a frequency response that is similar in shape, but $P$ times narrower than the frequency response of the original analysis filter.
A DFT applied to the larger number of samples will naturally produce a correspondingly larger number of (more closely spaced) frequency channels, but choosing only every $P$th channel and discarding the rest (known as \textit{decimation}) ensures that the desired spectral resolution with $K$ channels is retained.
In this way, spectral leakage can be contained arbitrarily close to the ``correct'' channel by choosing a sufficiently high value of $P$.

The two-step algorithm described above (performing a windowed DFT on $N = KP$ samples and decimating the resulting spectrum) defines the PFB.
Formally, it is equivalent to Eq. \eqref{eqn:wdft}; however, in this context the term \textit{critically sampled} (i.e. $M = K$) implies that the $N$-length windows will now overlap.
The term ``polyphase'' derives from the fact that each block of $K = N/P$ samples (known as \textit{taps}) in $x[n]$ is included in multiple applications of the DFT, but appearing at a different relative phase in each case.

\begin{figure}[p]
    \centering
    (a) \\
    \includegraphics[width=\columnwidth]{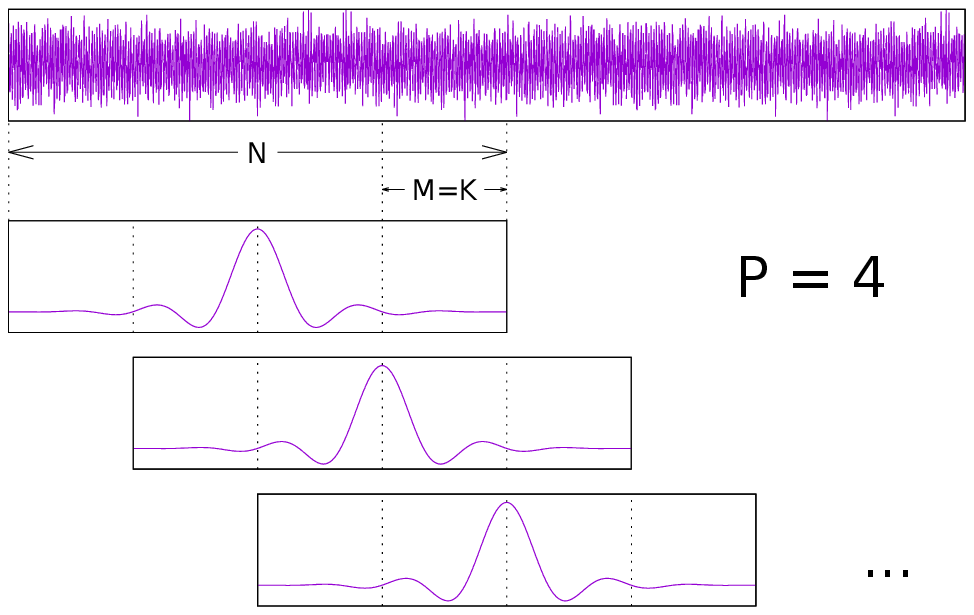} \\
    (b) \\
    \includegraphics[width=\columnwidth]{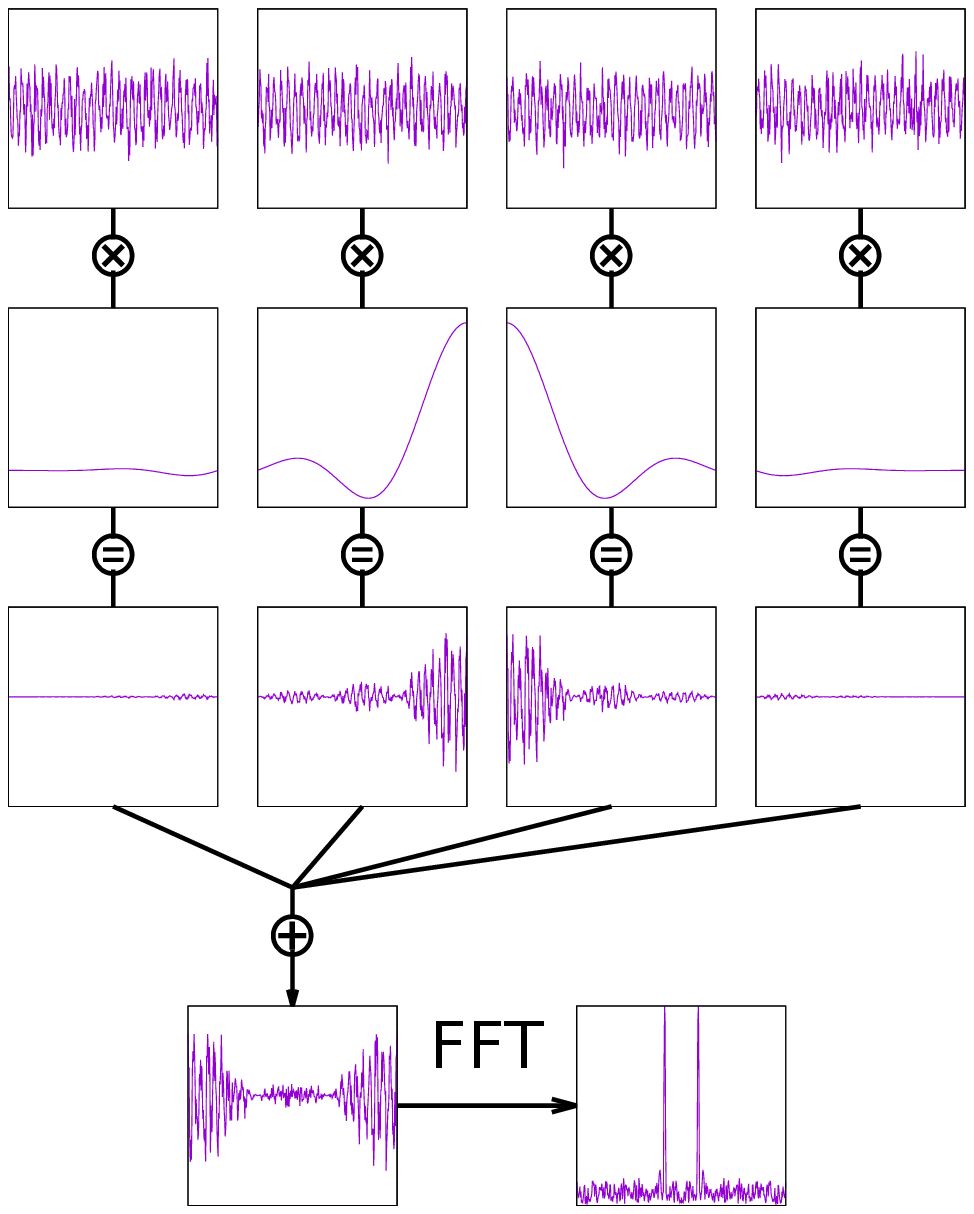}
    \caption{A diagrammatic representation of the weighted overlap-add algorithm, as defined in Eq. \eqref{eqn:pfb}. Panel (a) shows the filter window being translated along a discretely sampled signal (in this case, containing a sinusoid and noise) with a step size of one tap. At each step, panel (b) shows how the signal (first row) is multiplied by a filter (second and third rows), and each tap is summed (bottom left) and Fourier transformed to produce the final spectrum (bottom right).}
    \label{fig:pfb}
\end{figure}

One of the great advantages of the critically sampled PFB is the existence of a mathematically equivalent but computationally efficient implementation.
It can be shown that Eq. \eqref{eqn:wdft} is equivalent to first segmenting the windowed time series into taps, summing their respective samples element-wise, and performing a single DFT on the resulting array (now also of size $K$), i.e.
\begin{equation}
    X_k[m] = \sum_{n=0}^{K-1} b_m[n]\,e^{-2\pi jkn/K},
    \label{eqn:pfb}
\end{equation}
where
\begin{equation*}
    b_m[n] = \sum_{\rho=0}^{P-1} h[K\rho - n]\,x[n + mM - K\rho].
\end{equation*}
A short proof of this equivalence is given in \citet{Harris2011}.
The procedure described by Eq. \eqref{eqn:pfb} is called the \textit{weighted overlap-add} algorithm, and is illustrated in Fig. \ref{fig:pfb}.

% There is some ambiguity about whether this definition includes the correct set of phases, which differ depending on whether the filterbank is defined on [0,M) or [-M/2, M/2).
% It may turn out there is an extra phase ramp that's needed in Eq. \eqref{eqn:filter} below.
% See Harris et al. (2011) for a fuller discussion of this point.

\subsection{MWA implementation of the fine PFB}
\label{sec:mwa_pfb}

\begin{figure}[t]
    \centering
    \includegraphics[width=\columnwidth]{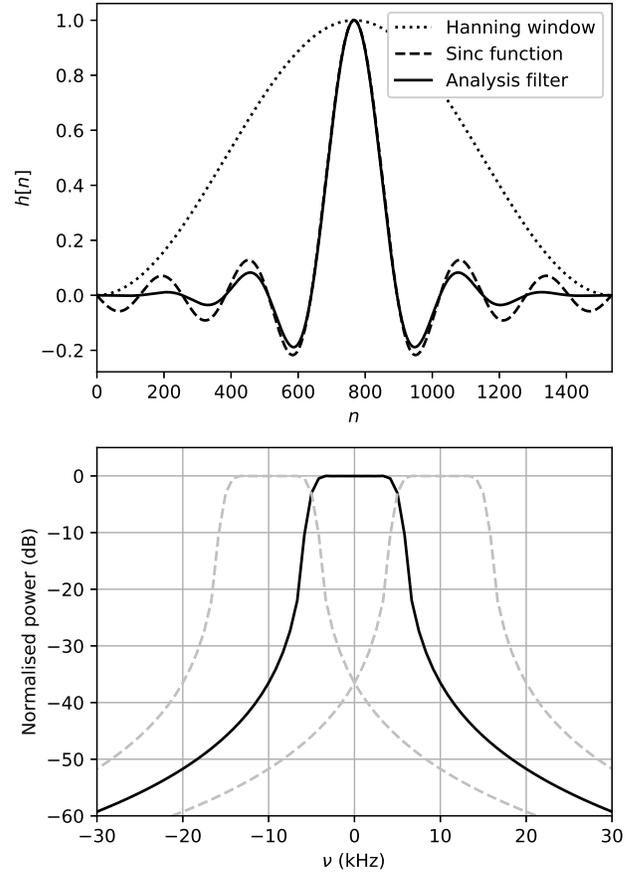}
    \caption{Top: The coefficients of the MWA's fine PFB analysis filter, defined in Eq. \eqref{eqn:filter}, which is composed of a Hanning window multiplied to a $\sinc$ function. Bottom: The frequency response of the analysis filter (black, solid), showing negligible attenuation across approximately $10\,$kHz (the bandwidth of a fine channel) and strong attenuation elsewhere. The frequency response is repeated for adjacent channels on either side (grey, dashed) showing crossover points on the channel edges at $-3\,$dB.}
    \label{fig:filter}
\end{figure}

The first stage (coarse) PFB is described in \citet{Prabu2015}, and here we only document a few details pertaining to the second stage (fine) PFB.
A PFB is specified by (1) the number of output channels, $K$, (2) the number of taps, $P$, and (3) the analysis filter, $h[n]$.
For the MWA, $K = 128$ (giving fine channels $10\,$kHz wide), $P = 12$, and
\begin{equation}
    h[n] =
        \begin{cases}
            w_H[n]\,w_s[n], & 0 \le n < N, \\
            0               & \text{otherwise}
        \end{cases}
    \label{eqn:filter}
\end{equation}
where
\begin{equation*}
    w_H[n] = \sin^2\bigg(\frac{\pi (n+1)}{N+1}\bigg)
\end{equation*}
is the Hanning window, and
\begin{equation*}
    w_s[n] = \sinc\bigg(\frac{\pi(n + 1 - N/2)}{K}\bigg)
\end{equation*}
is the scaled $\sinc(x) = \sin(x)/x$ function.
It can be easily checked that $h[n]$ is defined to be symmetric around sample $n = 767 = N/2-1$.
The analysis filter is shown in Fig. \ref{fig:filter}.
The MWA's fine PFB is critically sampled, with $M = K = 128$.
The rationale behind the particular design choices for the MWA's fine PFB is beyond the scope of this paper---for the purposes of creating a synthesis filter, it is sufficient to know merely how the analysis filter is defined\footnote{Eq. \eqref{eqn:filter} is defined on $\mathbb{R}$, but the actual implementation on the MWA's filed-programmable gate arrays (FPGAs) defines the analysis filter coefficients on $\mathbb{Z}$. The exact implemented values, $h^\ast[n]$, can be obtained from $h[n]$ via $h^\ast[n] \equiv \lfloor\alpha h[n]\rceil$, where $\lfloor\cdot\rceil$ denotes rounding to the nearest integer and $\alpha = 117963.875$.}, and the fact that the PFB is critically sampled.

The fine PFB is implemented on field-programmable gate arrays (FPGAs)\footnote{Xilinx Virtex4 XC4VSX35} and was designed in such a way to accommodate the data rate and bit depth constraints set by the surrounding hardware.
The input signal values are signed (5+5)-bit complex integers, and the final outputs are (4+4)-bit complex integers.
The loss of precision associated with the demotion of 1 bit naturally places limits on the ability for any synthesis filter to perfectly reconstruct the original coarse channel time series, which is analysed for our system below.
The full details of the FPGA implementation are given in the appendix.

\section{Synthesis filters}
\label{sec:ipfb}

In order to regain the time resolution lost during analysis, the spectral output of the PFB must be transformed back into the time domain.
The inverse to the analysis operation defined in Eq. \eqref{eqn:wdft} converts a spectrum $X_k[m]$ into a time series $\hat{x}[n]$ by means of a \textit{synthesis filter}, $f[n]$:
\begin{equation}
    \hat{x}[n] = \sum_{m = -\infty}^{\infty} f[n - mM]\,
        \frac{1}{K} \sum_{k=0}^{K-1} X_k[m]\,e^{2\pi jkn/K}.
    \label{eqn:ipfb}
\end{equation}
In this expression, the index $m$ is allowed to run over all integers in order accommodate an arbitrarily large synthesis filter.

Back-to-back analysis-synthesis filters can be designed so that the original time series can be perfectly reconstructed without any loss of signal (i.e. $\hat{x}[n] = x[n]$ exactly), and the system as a whole can be thought of as an identity operation.
The condition for perfect reconstruction can be found by substituting the analysed spectrum obtained from Eq. \eqref{eqn:wdft} into the synthesis operation defined in Eq. \eqref{eqn:ipfb}.
For a critically sampled system,
\begin{equation}
    \begin{aligned}
        \hat{x}[n]
            &= \sum_{s = -\infty}^{\infty} x[n - sM]
               \sum_{m = -\infty}^{\infty} f_m[n]\, h_{m-s}[-n] \\
            &= x[n] \sum_{m = -\infty}^{\infty} f_m[n]\, h_m[-n] +{} \\
            &\quad \sum_{\substack{s = -\infty \\ s \ne 0}}^{\infty} x[n - sM]
               \sum_{m = -\infty}^{\infty} f_m[n]\, h_{m-s}[-n]
    \end{aligned}
    \label{eqn:backtoback}
\end{equation}
where, following \citet{Crochiere1983}, we have adopted the shorthand notation
\begin{equation}
    \begin{aligned}
        h_\lambda[n] &\equiv h[n - \lambda M], \\
        f_\lambda[n] &\equiv f[n - \lambda M].
    \end{aligned}
    \notag
\end{equation}
Perfect reconstruction corresponds to the case when the only non-zero contribution comes from the $s = 0$ term, giving rise to the necessary and sufficient condition
\begin{equation}
    \sum_{m = -\infty}^{\infty} f_m[n]\,h_{m-s}[-n] = \Kron
    \label{eqn:invertibility_condition}
\end{equation}
for all values of $n$, where $\Kron$ is the Kronecker delta.

Finding the synthesis filter that perfectly inverts a given analysis filter is tantamount to solving Eq. \eqref{eqn:invertibility_condition} for $f[n]$.
However, exact solutions do not always exist, and in general, numerical methods must be employed to find a synthesis filter that minimises the reconstruction error.

\subsection{Reconstruction error}

In the context of an astrophysical signal, it is desirable to quantify the reconstruction error in terms of the signal-to-noise ratio (S/N).
If we assume that individual samples follow, and are dominated by, the same Gaussian noise statistics, then the reduction in S/N of a reconstructed sample can be estimated by considering the fraction of the power of the reconstructed sample that came from the original sample:
\begin{equation}
    [\text{S/N}]_\text{recon}
      = \frac{\left(\sum\limits_{m = -\infty}^{\infty} f_m[n]\,h_m[-n]\right)^2}{
            \sum\limits_{s = -\infty}^{\infty}
            \left(\sum\limits_{m = -\infty}^{\infty} f_m[n]\,h_{m-s}[-n]\right)^2
        }.
    \label{eqn:snr_long}
\end{equation}
This expression is invariant under the substitution $n \rightarrow n + \lambda M$, $\lambda \in \mathbb{Z}$, which indicates that the (average) reconstruction error is only a function of where the sample in question falls within a tap.
Consequently, any synthesis filter (except the exact inverse of the analysis filter, if it exists) will introduce a ``ringing'' effect into the reconstructed time series that has a period equal to the size of the PFB tap.

The ``leakage'' of power into other samples implied by Eq. \eqref{eqn:snr_long} can also manifest itself as spurious imaging of a ``true'' signal at intervals of one tap.
This can be seen by considering the effect of a single-sample impulse, with a power much greater than the ambient noise, on the reconstructed time series.
The sample itself, say, $x[n]$, will be reconstructed with high fidelity, as the reconstructed error is only comprised of contributions from the (relatively) low-level noise.
The reconstruction error of the sample $x[n+K]$, however, will be dominated by the term in Eq. \eqref{eqn:backtoback} involving $x[n]$, according to the relative weighting introduced by the analysis and synthesis filters.
Thus, the impulse will reappear at intervals of one tap, but where the relative power of each appearance is given by
\begin{equation}
    p(s) = \left(\sum\limits_{m = -\infty}^{\infty} f_m[n]\,h_{m-s}[-n]\right)^2.
    \label{eqn:ringing}
\end{equation}
This effect, termed temporal imaging, is discussed further in the context of specific filters.

\subsection{Optimal and sub-optimal filter designs}

Minimising the loss of S/N is equivalent to solving Eq. \eqref{eqn:invertibility_condition} using least squares regression.
Since any given reconstructed sample only receives contributions from samples spaced one tap apart, Eq. \eqref{eqn:invertibility_condition} can be thought of as $M$ independent conditions, one for each tap position, $n = 0, 1, 2, \dots, M-1$.
Thus, considering each tap position separately, each condition can be expressed as a minimal matrix equation,
\begin{equation}
    H^{(n)}F^{(n)} = D,
\end{equation}
where $H^{(n)}$, $F^{(n)}$, and $D$ are matrices whose elements are given by
\begin{equation}
    \begin{aligned}
        H^{(n)}_{ij} &= h_{P-1+j-i}[-n], \\
        F^{(n)}_j    &= f_j[n], \\
        D_i          &= \delta^{(P^\prime+P)/2}_i,
    \end{aligned}
\end{equation}
where $P$ is the number of taps in the analysis filter, and the size of $F^{(n)}$ is set to the desired number of (non-zero) taps in the synthesis filter, $P^\prime$.
The indices are chosen such that, owing to the finite size of the analysis filter, the smallest $H^{(n)}$ that captures every non-trivial term in Eq. \eqref{eqn:invertibility_condition} is the $(P^\prime+P-1) \times P^\prime$ matrix
\begin{equation}
    H^{(n)} =
    \begin{bmatrix}
       h_{P-1}[-n] & 0 & \cdots & 0 \\
       h_{P-2}[-n] & h_{P-1}[-n] & \cdots & 0 \\
       \vdots & \ddots & \ddots & \vdots \\
       0 & \cdots & h_0[-n] & h_1[-n] \\
       0 & \cdots & 0 & h_0[-n]
    \end{bmatrix}
\end{equation}
with row number $i = (P^\prime + P)/2$ (in $H^{(n)}$ and $D$) corresponding to the $s = 0$ term\footnote{Note that since $P = 12$ for the PFB considered here, only the case where $P^\prime$ is even has been considered.}.

Once the matrices $H^{(n)}$, $F^{(n)}$, and $D$ have been defined, solution by least squares regression can proceed in the usual way, yielding the best-fit filter coefficients
\begin{equation}
    \hat{F}^{(n)} = \left({H^{(n)}}^T H^{(n)}\right)^{-1} {H^{(n)}}^T D,
\end{equation}
where ${H^{(n)}}^T$ is the transpose of $H^{(n)}$.
With this notation, the reconstruction error can be more concisely expressed
\begin{equation}
    [\text{S/N}]_\text{recon}
      = \frac{(\hat{D}_{(P^\prime+P)/2})^2}{\hat{D}^T \hat{D}},
    \label{eqn:snr_short}
\end{equation}
where $\hat{D} = H^{(n)} \hat{F}^{(n)}$.

\begin{figure}
    \centering
    \includegraphics[width=\columnwidth]{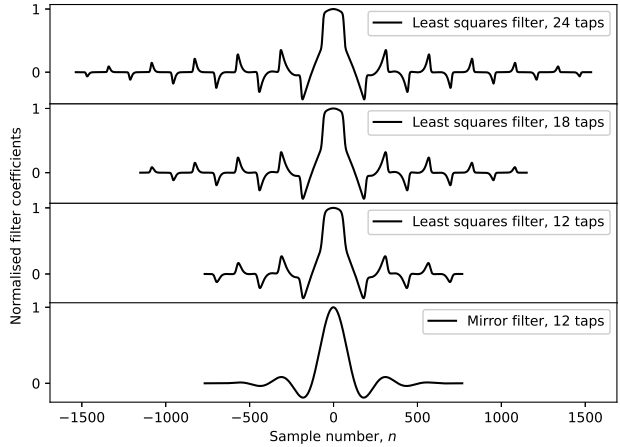}
    \caption{The coefficients for four different synthesis filter designs: three generated using least-squares optimisation methods, and the sub-optimal mirror filter.}
    \label{fig:synthfilters}
\end{figure}

Fig. \ref{fig:synthfilters} shows the solutions found for the MWA's analysis filter defined in Eq. \eqref{eqn:filter}, for 12-, 18-, and 24-tap synthesis filter sizes.
Despite the complexity of the synthesis filters, they all share the same basic form as the analysis filter.
This resemblance suggests that choosing a synthesis filter that is the mirror image of the analysis filter (i.e. $f[n] = h[-n]$), might also yield a reasonably good reconstruction, without having to calculate the optimal solutions numerically.

\begin{figure}
    \centering
    \includegraphics[width=\columnwidth]{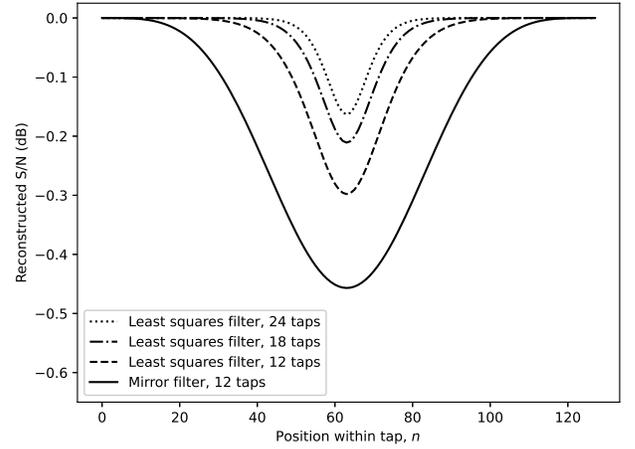} \\[10pt]
    \includegraphics[width=\columnwidth]{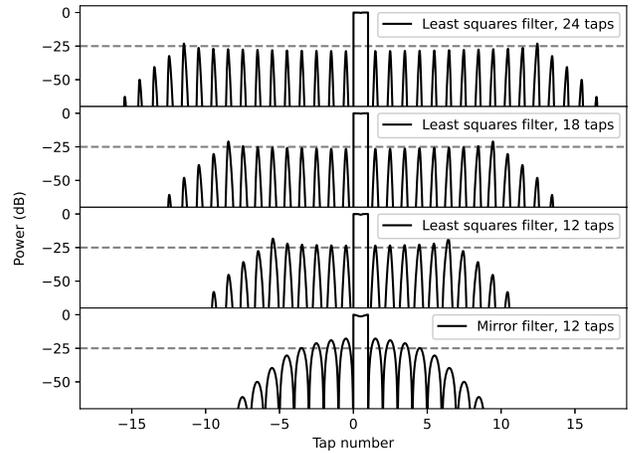}
    \caption{\textit{Top}: The reconstructed S/N, defined in Eqs. \eqref{eqn:snr_long} and \eqref{eqn:snr_short}, as a function of the position of the reconstructed sample within a tap for the filters displayed in Fig. \ref{fig:synthfilters}. \textit{Bottom}: The effect of temporal imaging, quantified in Eq. \eqref{eqn:ringing}, demonstrating how power ``leaks'' across adjacent taps during reconstruction. A grey dashed line is drawn at $-25\,$dB to aid visual comparison.}
    \label{fig:snr}
\end{figure}

The performance of both the (sub-optimal) mirror filter and the (optimal) set of least-squares solutions can be evaluated straightforwardly using Eq. \eqref{eqn:snr_short}.
Fig. \ref{fig:snr} compares the loss of S/N due to each filter (under the assumption of noise-dominated samples), revealing that, as expected, the filters with the larger number of taps perform better, and the mirror filter performs the least well.

Fig. \ref{fig:snr} suggests that one can achieve arbitrarily good performance by choosing a sufficiently long filter.
However, the better performance of the longer filters is offset by the increased computational resources and/or time required to apply them, which may exceed the system's design specifications.
Note that the mirror filter, being the same size as the 12-tap least squares filter, offers no advantage in terms of minimising the reconstruction error.
On the other hand, the way that the signal power is distributed across multiple taps is markedly different for the two filter types, as shown in the bottom panel of Fig. \ref{fig:snr}.
The mirror filter performs slightly worse in nearby taps, but drops off more quickly further out.
For applications where the samples are signal-dominated, and minimising temporal imaging is more important than measuring the S/N of the signal precisely, the mirror filter may therefore offer some advantage over the least squares solution.
However, observations of pulsars only rarely fall into this regime, even for observations of single pulses, where typically many samples must be averaged before the signal becomes more significant than the noise.
A more detailed analysis of these effects is therefore outside the scope of the present work.

\subsection{MWA Implementation}

We initially implemented the mirror filter for the purpose of prototyping the synthesis filter, due to the ease of generating the coefficients, and the 12-tap least squares filter was implemented once proof-of-concept had been demonstrated.
Both of these synthesis filters are available as optional components of the MWA-VCS beamforming software, \vcstools{}\footnote{\url{https://github.com/CIRA-Pulsars-and-Transients-Group/vcstools}}, described in \PaperI{}.
Even though our system places no stringent limits on computational resources, we have not implemented the larger filters due to the identification of other sources of error that dominate over the reconstruction error defined above, rendering the advantages of the longer filters relatively minor in comparison.
An analysis of these errors is presented in the following sections.

Viewed as a linear operation, Eq. \eqref{eqn:ipfb} is ideally suited for GPUs, and we have implemented it in \vcstools{} for NVIDIA's CUDA architecture.
Rather than use existing implementations of the implied inverse DFT, the $N \times K$ exponential terms (the so-called ``twiddle factors'') are pre-calculated and stored in a 2D array of double-precision floating point numbers on the GPUs.
These are then accessed as required for the calculation of a given sample $\hat{x}[n]$.
The combined GPU-accelerated calculations for both beamforming and coarse channel reconstruction run faster than real time\footnote{Verified for CUDA compute capability $>3.5$.}.

Once the synthesis filter has been applied to each of the MWA's polarisation streams for each coarse channel, the resulting high-time resolution time series is written out as complex voltages (i.e. without converting to Stokes parameters) in the VDIF file format \citep{Whitney2009}.
This format was chosen because it is supported by the coherent de-dispersion functionality of DSPSR\footnote{\url{http://dspsr.sourceforge.net/}} \citep{VanStraten2011b}, a software suite for processing pulsar time series.
The VDIF format requires each sample to fit into a signed 8-bit complex integer data type, which is performed as the final step before writing to disk.

\section{System performance}

The MWA implementation of the fine PFB differs from Eq. \eqref{eqn:pfb} in a few important respects, causing a reduction in the reconstructed S/N that would be present even if the synthesis filter perfectly inverted the analysis filter.
The most significant difference is that both the coefficients $b[n]$ and the final sum undergo a rounding operation, resulting in the approximate (i.e. both less precise and biased) spectrum
\begin{equation}
    \tilde{X}_k[m] = \left\lfloor\sum_{n=0}^{K-1} \bigg\lfloor b_m[n] \bigg\rfloor_{\text{asym}} e^{-2\pi jkn/K}\right\rfloor,
    \label{eqn:rounded}
\end{equation}
where $\lfloor\cdot\rfloor$ and $\lfloor\cdot\rfloor_{\text{asym}}$ denote symmetric and asymmetric rounding operations respectively, described in Appendix \ref{sec:fpga_pfb}.

The non-linearity of the back-to-back system induced by Eq. \eqref{eqn:rounded} implies that there is no single impulse response test that can adequately characterise the whole system.
To wit, the results of an impulse response test depend sensitively on at least three factors: the magnitude of the impulse; its position within a tap; and the arbitrary scaling factor and quantisation applied at the end of the test required by the integer-based output formats.
For example, an impulse at tap position $n \equiv 0$ (mod $K$) will be perfectly reproduced if its magnitude is such that the amount of rounding that takes place during analysis is minimised.
On the other hand, an impulse at $n \equiv 64$ (mod $K$) can be chosen such that the response is significantly worse than that implied by Fig. \ref{fig:snr}.

\begin{figure}
    \centering
    \includegraphics[width=\columnwidth]{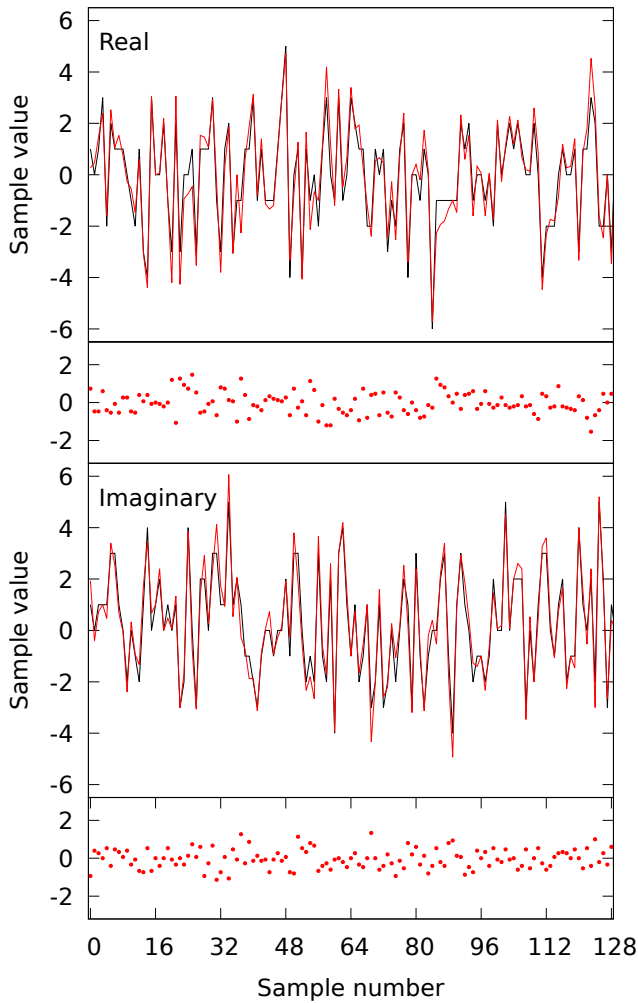}
    \caption{128 high time resolution ($1.28\,$MHz) samples from a single coarse channel and a single polarisation of one MWA tile. The real (top) and imaginary (bottom) components are shown separately. The original coarse channel data (black) are compared with the data reconstructed using the 12-tap least squares filter (red), with the residuals plotted in the lower panels. Because the VDIF data uses an arbitrary scaling factor, the red line has been rescaled by eye for a better visual comparison.}
    \label{fig:compare_timeseries}
\end{figure}

For this reason, and also because the vast majority of applications of this system falls in the regime of noise-dominated samples, we have decided to forego the traditional impulse response test in favour of a back-to-back test involving real data collected expressly for this purpose.
This is an empirical test of system that made use of a non-standard observing mode to record a small amount of simultaneous coarse and fine channel data (i.e. both before and after the fine PFB analysis filter stage).
The fine channels from one polarisation of a single tile were extracted and subjected to both the mirror filter and the 12-tap least squares filter to reconstruct the $1.28\,$MHz coarse channel time series, which could then be compared directly with the original data.
The real and imaginary parts of the time series resulting from the least squares filter are shown in Fig. \ref{fig:compare_timeseries}.

With the original and reconstructed time series in hand, we estimated the S/N loss for each tap position by comparing the variance of the original time series (the `signal'), $\sigma_S^2$, with the variance of the residuals (the `noise'), $\sigma_N^2$.
In analogy with Eq. \eqref{eqn:snr_long},
\begin{equation}
    [\text{S/N}]_\text{recon}
      = \frac{\sigma_S^2}{\sigma_S^2 + \sigma_N^2}.
    \label{eqn:snr_measured}
\end{equation}
Recognising that some of the noise variance is due to the synthesis filter (cf. Fig. \ref{fig:snr}), we have estimated the contribution due to rounding errors (and any other implementation-specific effects) by calculating the ``filter-subtracted'' S/N loss:
\begin{equation}
    [\text{S/N}]_\text{recon}
      = \frac{\sigma_S^2}{\sigma_S^2 + (\sigma_N^2 - \sigma_\text{filter}^2)},
    \label{eqn:snr_filter_subtracted}
\end{equation}
where $\sigma_\text{filter}$ is derived from Eq. \eqref{eqn:snr_long}.

\begin{figure}
    \centering
    \includegraphics[width=\columnwidth]{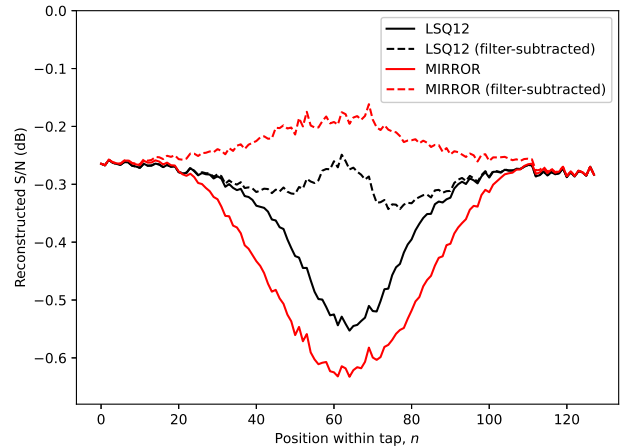}
    \caption{The S/N loss as a function of the position of the reconstructed sample within a tap for both the mirror filter and the 12-tap least squares filter. Both the total S/N loss (solid lines) and the S/N loss due to rounding and other implementation-specific effects (dashed lines) are plotted (cf. Eqs. \eqref{eqn:snr_measured} and \eqref{eqn:snr_filter_subtracted}).}
    \label{fig:snr_measured}
\end{figure}

Fig. \ref{fig:snr_measured} shows the results for just under one second's worth of data ($1278464 \times 0.78\,\mu$s samples), where 1536 samples (one tap) were excised due to the synthesis filter being applied to zero-padded data beyond the edge of the second.
The remaining samples were sorted into their respective tap positions, and the variances calculated for each set (containing $1278464/128 = 9988$ samples).
Near the edges of the tap, the rounding errors dominate the reconstruction noise, which we estimate contributes roughly $-0.26\,$dB of signal loss at all tap positions.
In more central tap positions, however, the contributions from the rounding errors and the filters are comparable, and significant improvements could made by using longer filters.
Nevertheless, unless there is a way to mitigate the rounding errors during the application of the analysis filter (explored below), it is unlikely that we will be able to achieve a total S/N loss better than approximately $-0.26\,$dB at any tap position.

\subsubsection*{Mitigation of rounding errors}

The original (i.e. before applying the fine PFB) high-time resolution samples are (5+5)-bit complex integers, scaled such that the bit occupancy is not so low that information is lost, and not so high that individual samples are clipped.
Interestingly, one may question whether or not this known quantisation can be used to recover some of the information lost by the imperfect back-to-back system.
If the errors are sufficiently small, then re-quantising the reconstructed samples will correct more errors than it introduces, giving overall better performance.

In this section, we offer a short proof of the condition for which re-quantisation results in a net recovery of S/N for samples following Gaussian statistics.
As before, let $\sigma_N^2$ be the variance of the residuals which are assumed to be drawn from a normal distribution with zero mean.
The effect of re-quantisation on the variance is that every sample between $k-\frac12$ and $k+\frac12$, for any integer $k$, gets ``counted'' as if it had the value $k$.
The adjusted variance, $\hat{\sigma}_N^2$, can then be expressed in terms of $\sigma_N^2$ via the second moment,
\begin{equation}
    \begin{aligned}
        \hat{\sigma}_N^2
         &= \sum_{k \in \mathbb{Z}} \int_{k-\frac12}^{k+\frac12} k^2
            \frac{1}{\sigma_N\sqrt{2\pi}} e^{-\frac{x^2}{2\sigma_N^2}}\,dx, \\
         &= \sum_{k \in \mathbb{Z}} k^2
            \left[\frac12\text{erf}\left(\frac{k+\frac12}{\sigma_N\sqrt{2}}\right) -
            \frac12\text{erf}\left(\frac{k-\frac12}{\sigma_N\sqrt{2}}\right)\right] \\
         &= \sum_{k=1}^{\infty} k^2
            \left[\text{erf}\left(\frac{k+\frac12}{\sigma_N\sqrt{2}}\right) -
            \text{erf}\left(\frac{k-\frac12}{\sigma_N\sqrt{2}}\right)\right],
    \end{aligned}
\end{equation}
where the last equality takes advantage of the symmetry of the error function.

Re-quantisation is beneficial precisely when $\hat{\sigma}_N^2 < \sigma_N^2$, which can be shown by numerical methods to occur when $\sigma_N^2 \lesssim 0.29$.
The MWA test data set presented above has a variance of approximately $6.7$, which implies that the total S/N loss of the back-to-back system would have to be smaller than $-0.05\,$dB before re-quantisation would improve the signal reconstruction.
As Fig. \ref{fig:snr_measured} shows, this is not met for any tap position, so we did not pursue this possibility further.
However, systems that are able to achieve better than $\sigma_N^2 \lesssim 0.29$ may do better yet by re-quantising the suitably scaled reconstructed samples.

\subsection{Verification using pulsar observations}
\label{sec:pulsardata}

Observations of three pulsars are presented here to validate the synthesis filter as a useful tool for undertaking high time resolution studies of pulsars with the MWA: MSPs \psrkaurJ{} and \psrbhatJ{}, and the bright, long-period PSR \psrslowB{}.
The first two pulsars were processed with the mirror filter, and \psrslowB{} was processed with the 12-tap least squares filter.
As Fig. \ref{fig:snr_measured} implies, the use of the mirror filter instead of the more optimal least squares filter would result in a further $\sim 0.1\,$dB reduction in S/N.

\subsubsection{PSR \psrkaurJ{}}

\begin{figure}[t!]
    \centering
    \includegraphics[angle=270,width=\columnwidth]{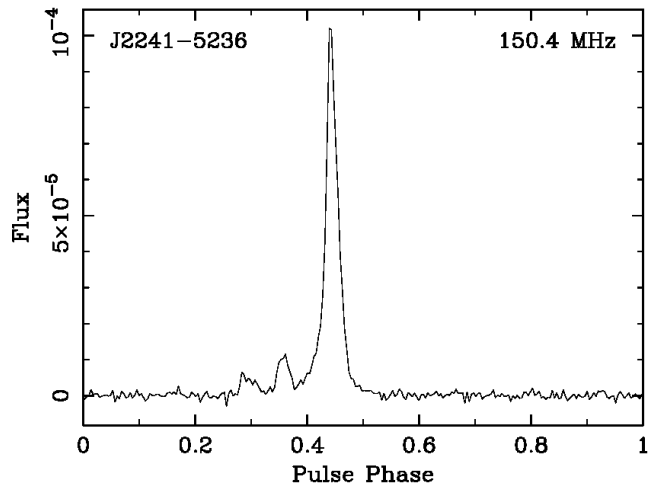}
    \caption{The coherently de-dispersed profile of PSR \psrkaurJ{} made from the reconstructed coarse channels \citep{Kaur2019} (the flux scales are arbitrary). The two pre-cursor components on the leading side of the main pulse are separated by approximately $50\,\mu$s, and were first detected using the high time resolution mode.}
    \label{fig:J2241_profiles}
\end{figure}

\psrkaurJ{} has a rotation period of $2.18\,$ms and a dispersion measure (DM) of $11.41\,\dmunits$.
The resulting dispersion smear across $10\,$kHz channels at the central observing frequency of $150.4\,$MHz is approximately $0.2\,$ms, or $45^\circ$ of rotation phase, consistent with the width of the average profile formed from incoherently de-dispersed, fine channelised data recorded with the MWA-VCS system.
Upon applying the synthesis filter and coherently de-dispersing the reconstructed time series sampled at the much higher rate of $1.28\,$MHz, the same data revealed exquisite detail in the average profile, including a pair of pre-cursor components that are only marginally visible (if at all) at higher frequencies (Fig. \ref{fig:J2241_profiles}).
These have since been confirmed during follow-up observations (Kaur et al., in prep) using the Band 3 receiver ($250$ to $500\,$MHz) of the upgraded Giant Metrewave Radio Telescope (uGMRT).
In addition, these new, low-frequency observations have allowed us to measure the DM of this pulsar with unprecedented precision ($\sim (2$-$6) \times 10^{-6}\,\dmunits$), with important consequences for measuring DM chromaticity and evaluating its effect on pulsar timing experiments.
These results are discussed in detail in \citet{Kaur2019}.

%The resemblance of the newly-sighted precursor components to the temporal artefacts induced by the synthesis filter naturally raises questions about their authenticity.
%The fact that the components only appear on the leading side of the main pulse is a point in favour of them being real, since the temporal artefacts are expected to appear on both leading and trailing sides.
%In addition, the time interval between the main pulse and the closest component is $\sim150\,\mu$s, which, although it corresponds to the expected location of the third-order artefact, is not expected to be brighter than the first two (see Fig. \ref{fig:impulse_response}); however, no evidence of the lower-order artefacts is seen.
%Furthermore, our recent (independent) follow-up observations using the Band 3 receiver ($250$ to $500\,$MHz) of the upgraded Giant Metrewave Radio Telescope (uGMRT) have confirmed that at least the innermost component is real (Kaur et al., in prep).

\subsubsection{PSR \psrbhatJ{}}

\begin{figure}[t!]
    \centering
    \includegraphics[angle=270,width=\columnwidth]{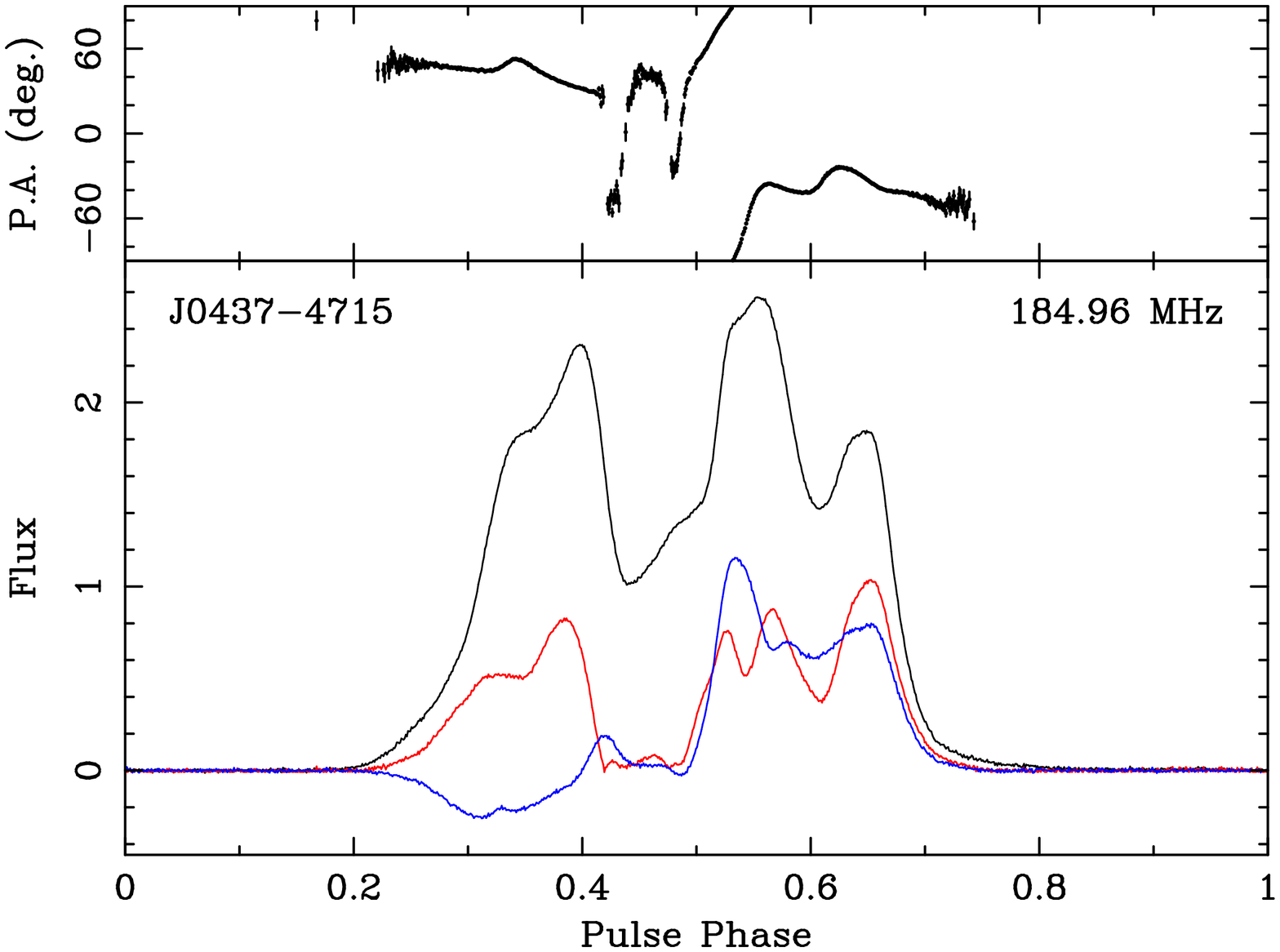} \\[10pt]
    \includegraphics[angle=270,width=\columnwidth]{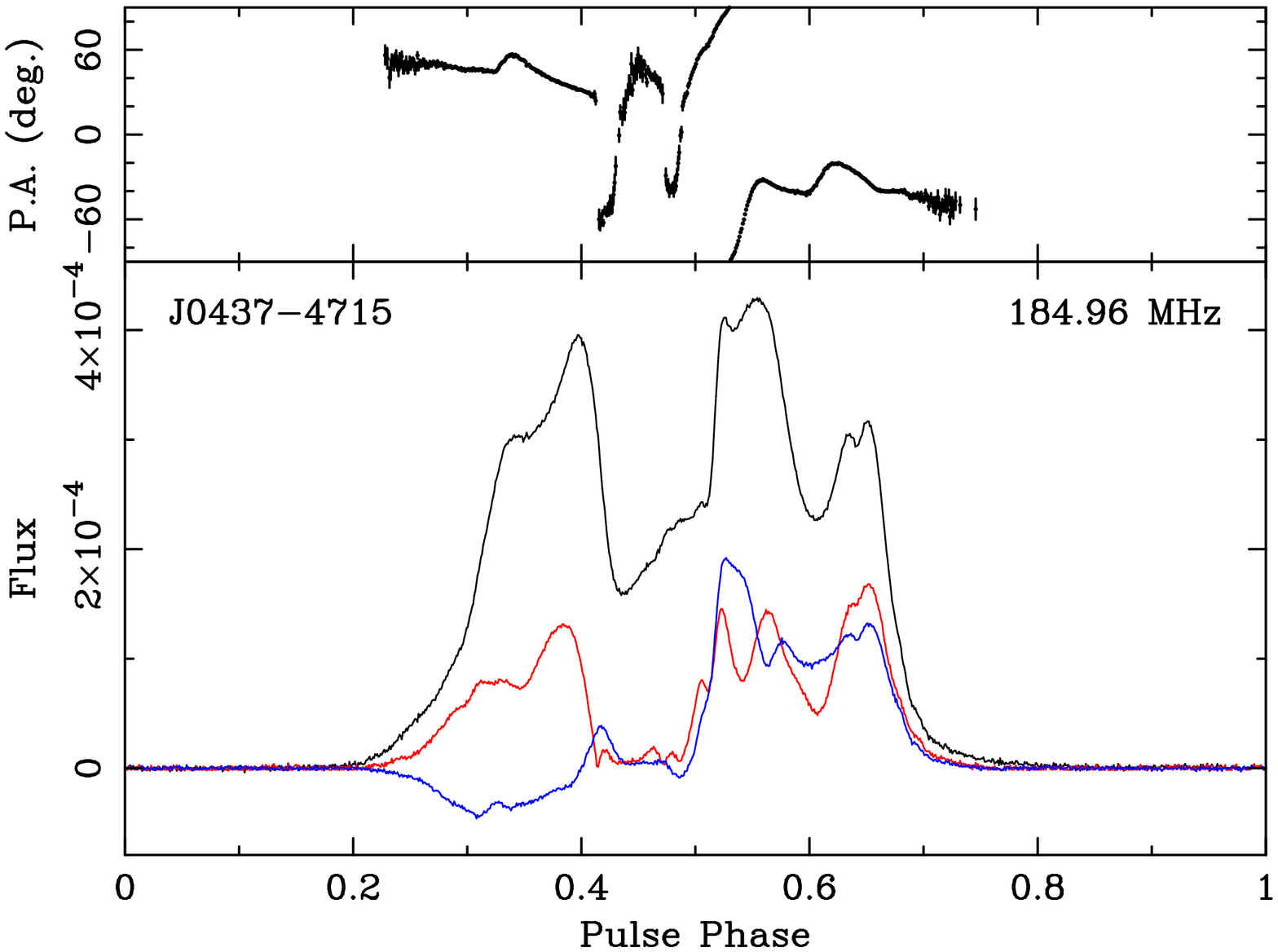}
    \caption{Profiles of PSR \psrbhatJ{} formed from the same data set. Top: incoherently de-dispersed with $10\,$kHz (fine) frequency channels formed from the standard PFB analysis filter. Bottom: coherently de-dispersed using $1.28\,$MHz (coarse) frequency channels reconstructed using the synthesis filter described in this paper. The higher time resolution profile shows features (e.g. notches in the total intensity profile at pulse phases $\sim 0.54$ and $\sim 0.7$) that are obscured by dispersion smear at the lower time resolution. A comparison with these profiles with that published in Paper I reveals an excess of circular polarisation on both the leading and trailing edges of the profile. The reason for the discrepancy is not clear, but it should be noted that the earlier profiles were published before the polarimetric verification of Paper II was carried out.}
    \label{fig:0437}
\end{figure}

PSR \psrbhatJ{} is a nearby MSP (distance $\approx 157\,$pc; period $P \approx 5.76\,$ms) and an established target for pulsar timing array (PTA) applications.
It boasts a complex, multi-component polarimetric profile (Fig. \ref{fig:0437}) that spans more than half of its rotation period \citep[cf.][]{Yan2011}.
It was used in Paper I as a verification of the beamforming method employed in the VCS processing pipeline, although there is a noticeable difference between the circular polarisation published in that work and that shown here (see the Figure caption for details).

As evident from Fig. \ref{fig:0437}, the application of the synthesis filter successfully recovers some curious fine structure (e.g. notch-like features at pulse phases $\sim$ 0.54 and $\sim$ 0.7) characteristic to this pulsar, first reported in early high time resolution studies made at 430 MHz \citep{Navarro1997}. The pulsar detection (for Stokes I only) was originally presented in \citet{Bhat2018}, where the combination of a lower time resolution (100 $\mu$s) and a non-negligible dispersive smearing ($\sim$45 $\mu$s) obscured the detection of these fine structures. Fig. \ref{fig:0437} thus presents the highest-fidelity detection of this important pulsar at frequencies below 200 MHz.

%\dots

\subsubsection{PSR \psrslowB{}}

\begin{figure}[t!]
    \centering
    \includegraphics[width=\columnwidth]{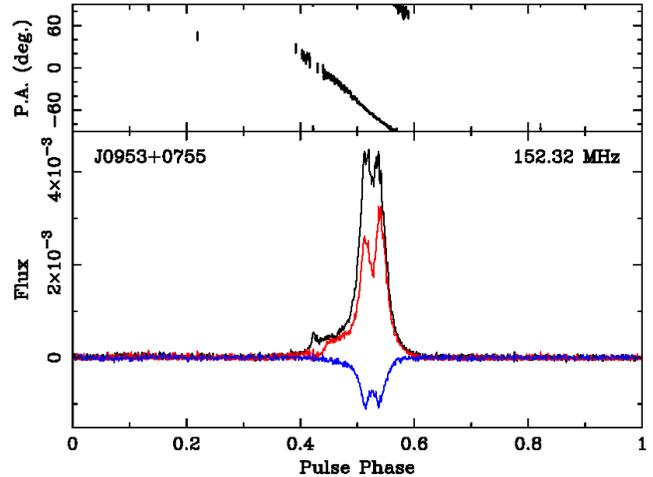}
    \caption{Polarisation profile formed using 80 seconds (315 pulses) of \psrslowB{} data. The time resolution is $\sim 250\,\mu$s.}
    \label{fig:0950_80_secs}
\end{figure}

\begin{figure}[t!]
    \centering
    \includegraphics[width=0.7\columnwidth,angle=270]{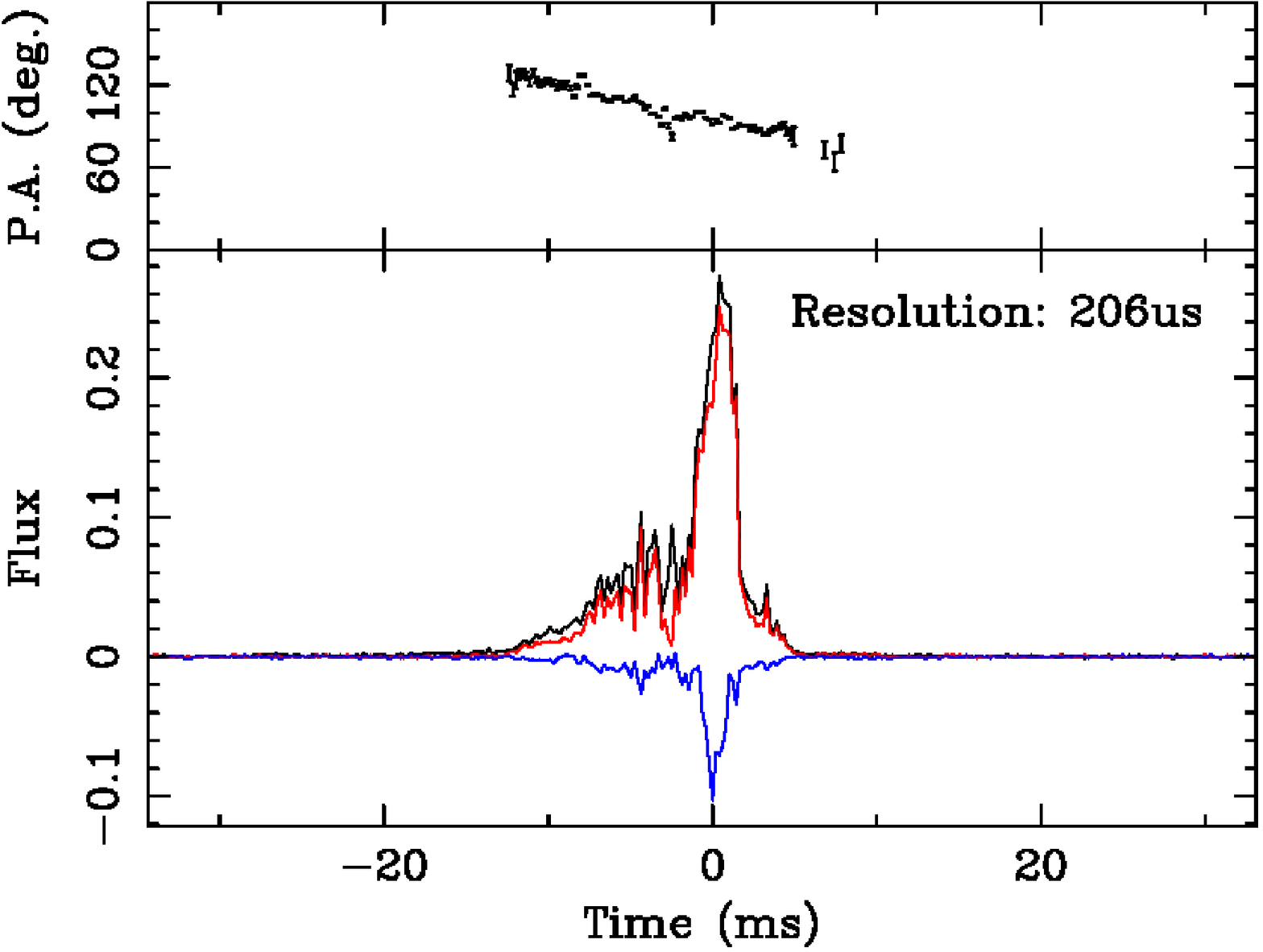} \\[10pt]
    \includegraphics[width=0.7\columnwidth,angle=270]{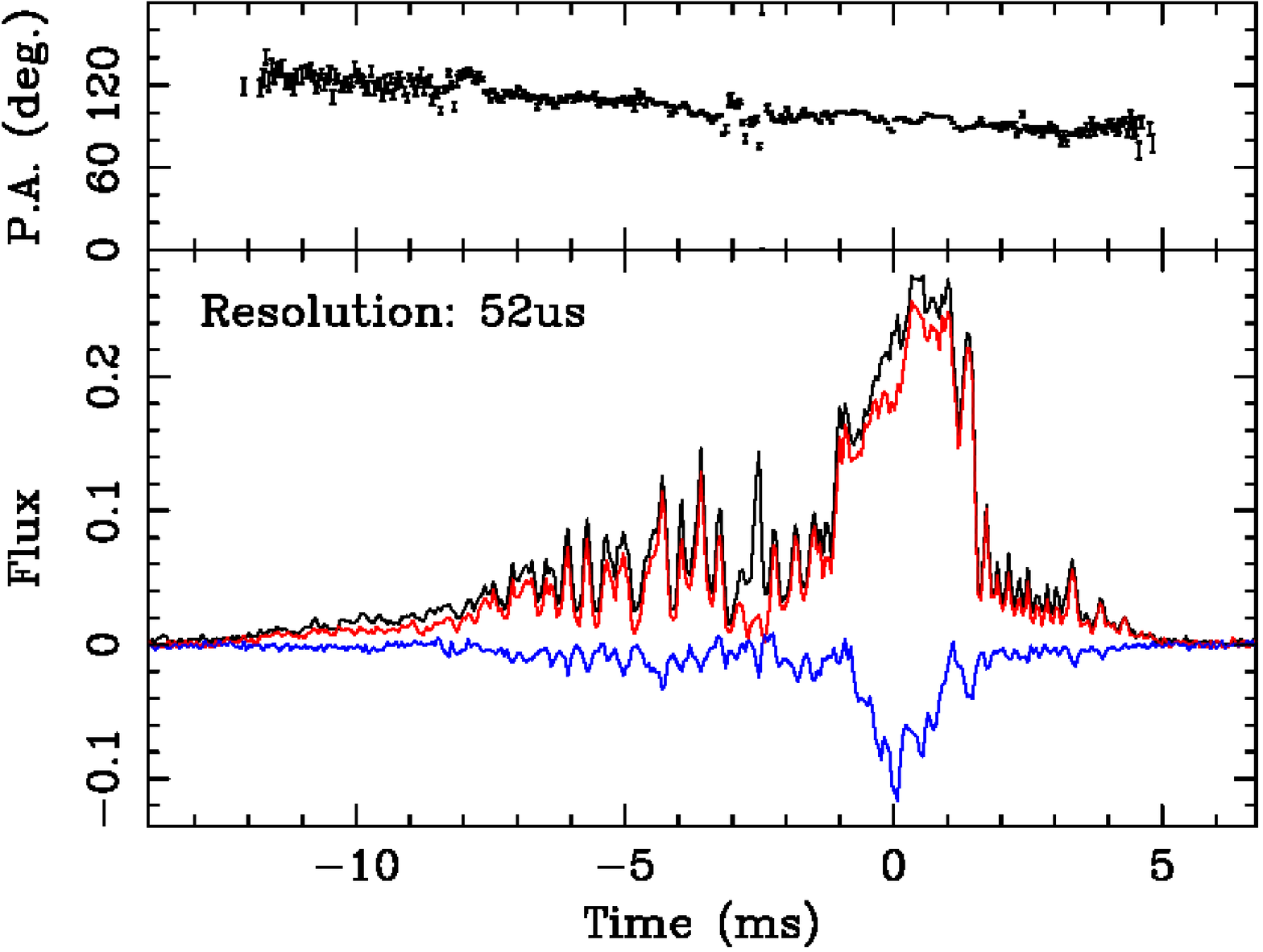} \\[10pt]
    \includegraphics[width=0.7\columnwidth,angle=270]{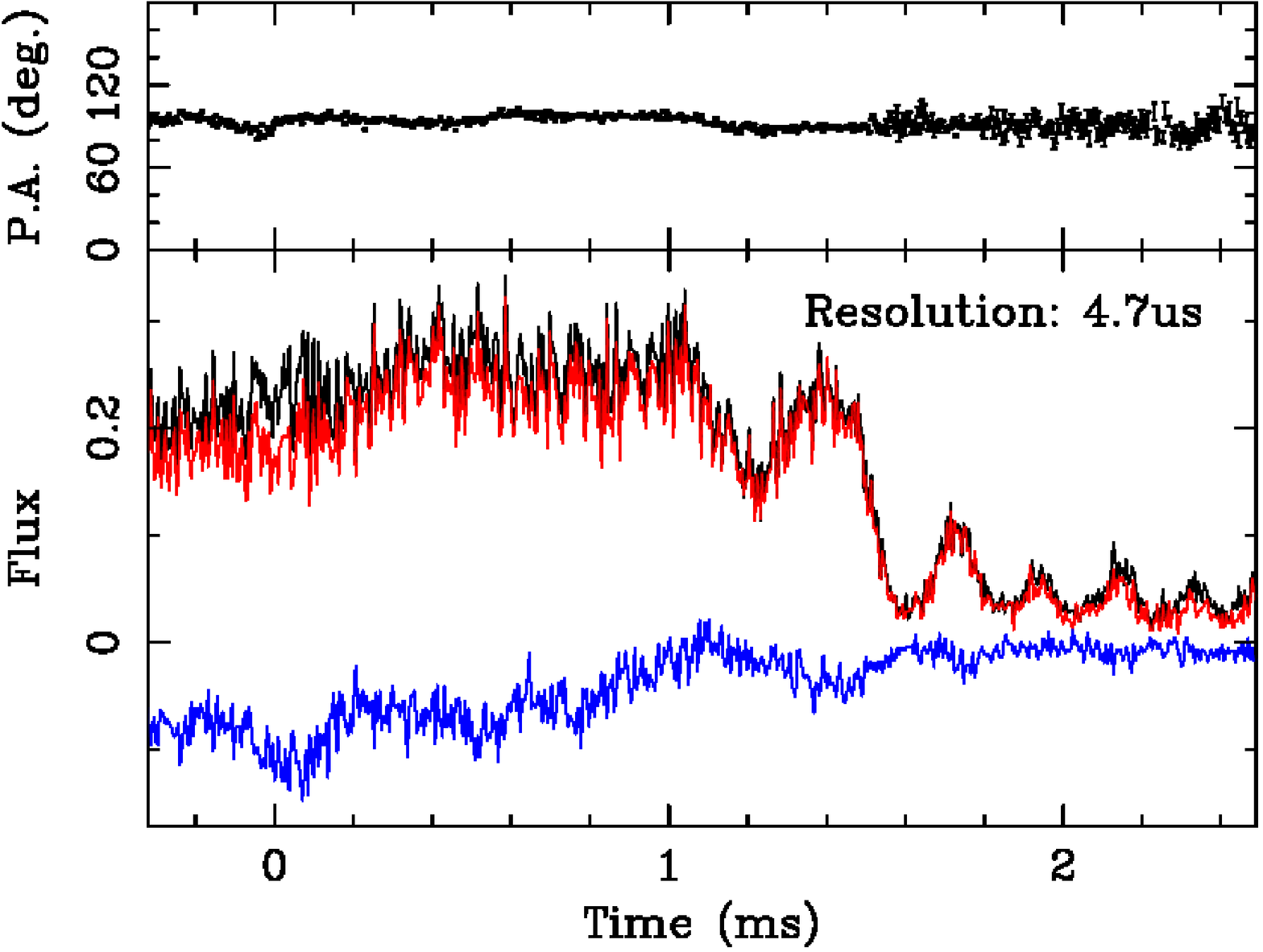}
    \caption{A bright, single pulse of \psrslowB{}, integrated over $24 \times 1.28\,$MHz coarse channels, and displayed with full polarisation at time resolutions of $\sim 200\,\mu$s (top), $\sim 50\,\mu$s (middle), and $\sim 5\,\mu$s (bottom).}
    \label{fig:0950_single_pulse}
\end{figure}

\psrslowB{} is a bright, long-period ($P = 0.253\,$s) pulsar known to exhibit microstructure \citep{Popov2002,Kuzmin2003}.
A total of eighty seconds (315 pulses) of data ($128 \times 10\,$kHz fine channels) were recorded, beamformed, and subjected to the polyphase synthesis filter.
Across the MWA bandwidth, we used the RM synthesis technique\footnote{\url{https://github.com/gheald/RMtoolkit}} \citep{Brentjens2005} to measure a rotation measure (RM) of $1.43 \pm 0.15\,$rad m$^{-2}$, which is consistent with previous measurements of this pulsar's RM \citep[e.g.][]{Noutsos2015}.
The mean profile formed from the one coarse channel is shown in Fig. \ref{fig:0950_80_secs}.
Several individual pulses were sufficiently bright to see substructure.
One such pulse is showcased in Fig. \ref{fig:0950_single_pulse}, where all 24 coarse channels were integrated to maximise the S/N.
To highlight the pulse's substructure, we show it at three different time resolutions (corresponding to the chosen bin width of each plot).

Using observations at $102.5\,$MHz, \citet{Kuzmin2003} report microstructure with a characteristic timescale of $\sim 60\mu$s, which we are able to unambiguously identify in our data.
We suggest that the finest observable structures (Fig. \ref{fig:0950_single_pulse}, bottom plot) might correspond to the $\sim 10\,\mu$s structures reported by \citet{Popov2002} at $1.65\,$GHz.
However, \citet{Kuzmin2003} used only a single linear polarisation feed, and \citet{Popov2002}, only a left-handed circularly polarised feed.
With the benefit of full Stokes polarimetry, we are able to verify that on the smallest time scales, the emission is almost 100\% polarised---mostly linear, but with a small amount of circular polarisation during the brightest part of the pulse.
This is evidence that the errors introduced by the analysis-synthesis filter do not contribute a significant amount of polarisation leakage into neighbouring time bins---in particular, at $50\,\mu$s intervals.

\section{Conclusion}

The MWA-VCS telescope system outputs voltage data with a frequency resolution of $10\,$kHz and a time resolution of $100\,\mu$s, which is suitable for many pulsar applications, as demonstrated in several pulsar science papers published to date.
In this paper, we have presented an algorithm to undo the fine channelisation stage (fine PFB) of the VCS pipeline by using a synthesis polyphase filter, implemented as part of our processing software suite, \vcstools{}.
The result is a reconstructed, coarse channelised (pre-PFB) time series, suitable for high-time resolution studies of MSPs and other rapid transient signals.
Although the reconstruction is not perfect, the average error does not exceed $-0.65\,$dB for noise-dominated samples.

We have further verified our system by observing PSRs \psrkaurJ{}, \psrbhatJ{}, and \psrslowB{}, all of which are known to exhibit structures on $\sim 1$-$10\,\mu$s timescales.
In each case, our results have been shown to be consistent with observations at other radio frequencies.
The advent of the VCS high time resolution observing mode thus distinguishes the MWA as a premier low-frequency instrument for studying pulsars and other transients at microsecond resolution, with important implications for studies of the pulsar emission mechanism, the characterisation of the ISM in the ongoing search for gravitational waves, and others.

\begin{acknowledgements}
We thank the anonymous referee for suggestions that greatly improved this work.
This scientific work uses data obtained from the Murchison Radio-astronomy Observatory. We acknowledge the Wajarri Yamatji people as the traditional owners of the Observatory site.
Support for the operation of the MWA is provided by the Australian Government (NCRIS), under a contract with Curtin University administered by Astronomy Australia Limited.
SM acknowledges the contribution of an Australian Government Research Training Program Scholarship in supporting this research.
DK acknowledges support from both the Curtin International Postgraduate Research Scheme and the NANOGrav project, which receives support from NSF Physics Frontier Center award number 1430284.
This work was further supported by resources provided by the Pawsey Supercomputing Centre with funding from the Australian Government and the Government of Western Australia.
We would also like to thank Ian Morrison, Franz Kirsten, and Clancy James for reviewing this manuscript and offering several helpful suggestions to improve it.
\end{acknowledgements}

\begin{appendix}

\section{The FPGA implementation of the fine PFB}
\label{sec:fpga_pfb}

Each coarse channel is independently processed by dedicated FPGAs to convert (5+5)-bit complex integers sampled at $1.28\,$MHz into a series of spectra composed of $128\times$(4+4)-bit complex integers sampled at $10\,$kHz.
The PFB is implemented in two stages: (1) a ``front-end'' stage which prepares the array $b[n]$ (see Eq. \eqref{eqn:pfb}), and (2) the ``FFT'' stage, which calculates the 128-point spectrum, $X_k[m]$.

\begin{figure}[t!]
    \centering
    \includegraphics[width=\columnwidth]{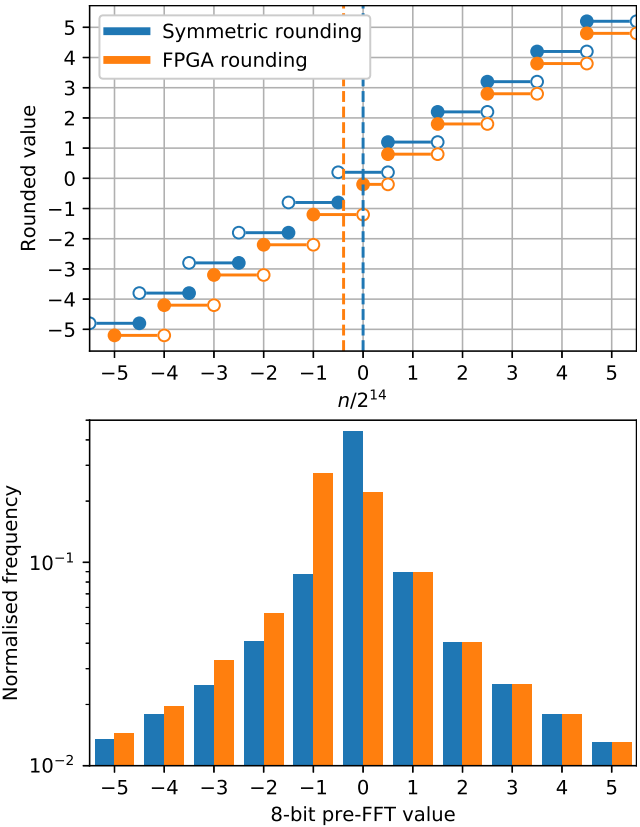}
    \caption{The first-stage (i.e. pre-FFT) rounding scheme implemented in the MWA FPGAs compared with a symmetric scheme. The top panel illustrates the two rounding schemes (each has been given a slightly different $y$-offset for visual clarity). A population of 5-bit samples was drawn from a rounded Gaussian distribution with mean $\mu = 0$ and standard deviation $\sigma = 2$. The vertical dashed lines show the means of the distributions after the population has been subjected to the first stage of the PFB (i.e. converted to 8-bit integers) and rounded according to the two different schemes. The bottom panel compares the distributions of the resulting population of 8-bit samples.}
    \label{fig:pre_fft_distributions}
\end{figure}

The first stage involves a multiplication of a window of 1536 consecutive coarse channel samples with the analysis filter shown in Fig. \ref{fig:filter}, and then a summation of the 12 taps together to produce a single array of 128 complex integers.
The combination of the allowed input values $[-16, 15]$ and the filter values guarantees that the magnitudes of the summed (signed) numbers do not exceed $2^{21} = 2097152$.
At this stage of the processing, they are stored as 48-bit signed integers, of which only the bottom 22 bits are therefore significant.
Each integer $n$ (either real or imaginary) is then reduced from 48 bits to 8 bits in the following manner.
If $n$ is positive, then bits 14 through 21 (counting from the least significant bit) are selected to form the 8-bit integer, and 1 is subsequently added if bit 13 is 1.
This is equivalent to rounding the number $n/2^{14}$ to the nearest integer, where fractional values of 0.5 are always rounded up.
If $n$ is negative, then bits 14 through 21 are selected, but no rounding occurs.
This is equivalent to applying the floor operation to $n/2^{14}$.
It should be noted that this rounding scheme introduces a bias into the distribution of values.
In particular, the distribution of 8-bit values has a different mean than the distribution of the original 5-bit values, and it results in an artificial deficit in the number of values that get rounded to zero.
The rounding scheme and its effect on the distribution are illustrated in Fig. \ref{fig:pre_fft_distributions}, including a displaced mean which artificially adds power to the DC bin of the spectrum.

The second stage calculates the spectrum of $b[n]$ using the standard fixed point FFT algorithm that is implemented on (Xilinx Virtex4 XC4VSX35) FPGAs.
The output values are scaled and (symmetrically) rounded, producing (4+4)-bit output values.

\end{appendix}

\bibliographystyle{pasa-mnras}
\bibliography{biblio.bib}

\end{document}